\documentclass[fleqn,10pt]{SelfArx} 

\definecolor{color1}{RGB}{0,0,90} 
\definecolor{color2}{RGB}{0,20,20} 

\newlength{\tocsep} 
\usepackage{lipsum} 
\usepackage{amsmath}
\usepackage[english]{babel}
\usepackage{graphicx}
\usepackage{multicol}
\usepackage{subfigure}
\usepackage[numbers]{natbib}
\usepackage{color}
\usepackage{epstopdf} 
\usepackage{multirow}

\JournalInfo{Submitted to CSAR2013 } 
\Archive{Preprint} 

\PaperTitle{Application of a cognitive-inspired algorithm for detecting communities in mobility networks} 

\Authors{Emanuele Massaro\textsuperscript{1}*, Lorenzo Valerio\textsuperscript{2}, Andrea Guazzini\textsuperscript{3}, Andrea Passerella\textsuperscript{2} \\and Franco Bagnoli\textsuperscript{4}} 
\affiliation{\textsuperscript{1}\textit{ Department of Information Engineering, University of Florence, Italy}} 
\affiliation{\textsuperscript{2}\textit{Institute for Informatics and Telematics, National Research Council, Pisa, Italy}} 
\affiliation{\textsuperscript{3}\textit{Department of Education and Psychology, University of Florence, Italy}}
\affiliation{\textsuperscript{4}\textit{Department of Physics and Astronomy, 
    University of Florence, and INFN Italy}}
\affiliation{*\textbf{Corresponding author}: emanuele.massaro@unifi.it} 

\newcommand{\keywordname}{Keywords} 

\Abstract{The emergence and the global adaptation of mobile devices has influenced human interactions at the individual,
community, and social levels leading to the so called Cyber-Physical World (CPW) convergence scenario~\cite{Conti20122}. One of the most
important features of CPW is the possibility of exploiting information about the structure of the social communities of users, revealed by joint movement patterns and frequency of physical co-location.  Mobile devices of users that belong to the same social community are likely to "see" each other (and thus be able to communicate through ad-hoc networking techniques) more frequently and regularly than devices outside  the community. In mobile opportunistic networks, this fact can be exploited, for example, to optimize networking operations such as forwarding and dissemination of messages.  In this paper we present the application of a cognitive-inspired algorithm~\cite{Massaro2012, Bagnoli2012, Guazzini2012} for revealing the structure of these dynamic social networks (simulated by the HCMM model~\cite{Boldrini:2010fk}) using information about physical encounters logged by the users' mobile devices. 
The main features of our algorithm are: (i) the capacity of detecting social communities induced by physical co-location of users through distributed  algorithms; (ii) the capacity to detect users belonging to more communities (thus acting as bridges across them), and (iii) the capacity to detect the time evolution of communities.}

\begin{document}

\flushbottom 

\maketitle 

\tableofcontents 

\thispagestyle{empty} 


\section{Introduction} 
Nowadays, the closer and closer interaction between devices and their users is a clear expression of the increasing tightness among  the cyber world and the physical one. Let us consider, for example, mobile devices that are in charge of autonomously accomplishing tasks like that of discerning, collecting and redistributing  important information (for their users) that can be collected in the environment. On the one hand, the devices can use  the information coming from the physical world to adapt and optimize their behaviour in the cyber world and, on the other hand, the feedback of the mobile device in the cyber world can affect the behaviour of their human users in the physical world (as happens in social gaming or with other social-oriented applications). 
This strong interaction has not only the quite obvious effect of generating a huge amount of information that flow from one world to the other, but it also triggers  a deeper connection between the them, leading to the so called Cyber-Physical World (CPW)  convergence scenario \cite{Conti20122}.  In this context,  mobile devices play an important role because they are the actual representation of their users in the cyber world or in other terms, mobile devices act as proxies of their human counterparts.  
 The challenge here is to devise methodologies that make devices able to properly mine the acquired knowledge in order to make them aware about their environment so that they can autonomously take proper decisions for specific tasks. Opportunistic Networks (OppNets) and the problems connected to them, represent a  perfect example of the this general concept.  OppNets~\cite{Pelusi2006} are dynamic, delay-tolerant wireless networks made by mobile nodes (\textit{e.g.} human users equipped with smartphones) where the connectivity between them is not guaranteed at any time instant. In OppNets the communication between nodes can occur only upon contacts, (\textit{i.e.} when nodes are in a reciprocal transmission range) and the information spreading mainly occour through  the \emph{store carry and forward} paradigm: nodes exploit any contact with other peers to exchange messages under the condition that the other peer is deemed  a good candidate to bring the message closer to the destination. The efficient delivery  of information to  interested users in this kind of networks is currently an open research problem. To this goal, researchers not only have to consider the typical physical problems of wireless networks but also the aspects connected with the humans' behaviour like their mobility patterns, their natural tendency to aggregate in social communities, etc. The ability of catching and understanding such social information, in order to predict and exploit human behaviour, has a great relevance for the development of effective solution for the above mentioned problems in OppNets. Let us consider for example the message forwarding problem in OppNets: due to the high mobility of devices, the challenge for a forwarding method is to quickly forward the message from the source to the destination, without introducing too many duplicate messages or overhead information. Here, the nodes' awareness about information like the social relationships,  the aggregation habits and the community structure of their human users (all information coming from the physical world and exploited in the cyber world), can help to select suitable forwarders while containing the delivery costs. In this work we focus on the community detection problem in occasional co-located mobile agents.  In other terms  we want to identify in real-time and using a distributed algorithm the  dynamical network structure emerging by proximity contacts of mobile agents.  The idea is that these device should be able to detect, in a dynamical and decentralized way,   the community structure  their users happen to belong to. We recall that in our scenario nodes must be able to take proper decision without relying on centralised information so it is very important that nodes autonomously build a local representation of their surrounding environment. Many community detection algorithms are presend in the literature, as reported in Ref.~\cite{Fortunato}. Many well-performing  algorithms for detecting communities in complex networks have been presented in the last decade.  We refer among the others to the so-called OSLOM~\cite{oslom}, INFOMAP and HIERARCHICAL INFOMAP~\cite{infomap, hinfomap}, MODULARITY OPTIMIZATION~\cite{mopt}, LOUVAIN METHOD~\cite{louvain} and the LABEL PROPAGATION METHOD~\cite{lprop} . Although they are very useful for offline data analysis on mobility traces and to define at priori strategies of data forwarding, data dissemination, energy saving, etc., they are rather unfit for real-time distributed applications,  \textit{i.e.}, for distributed algorithms run by mobile devices.  There are also centralized algorithm that can be applied to dynamic networks~\cite{survey, maria} or distributed ones that use global information~\cite{Hui, williams}. We assume here that the mobile device have no access to global data or global communication. 
Several decentralised approaches have been proposed for community detection. Differently from the centralized ones, they do not rely on a global vision of the network but only on a local one,\textit{ i.e.},  every node in the network builds and updates its own representation of the existing social communities over time. For example, in Ref.~\cite{Hui:2007uq,Hossmann2010}, the authors presented three community detection algorithms (SIMPLE, k-CLIQUE, and MODULARITY) while another improved one can be found in the work by Borgia et al.~\cite{Borgia:2011kx}. All these methods use only  the contact duration to build the representation of the social structure. Another important class of community detection algorithms are based on the local representation of the community,  as reported for example in Refs.~\cite{clauset, luo}.

We tackle the problem from a different  point of view, considering also some social and psychological aspects of human behaviour.  Human communities are large and varied; we recognize several levels of grouping, sometimes dependent on the context, and we have probably developed our language as a tool for faster communication and discovering of social relationships. Therefore in social networks it is very difficult to have a precise definition of community because people often belong to different communities at the same time and there is not a clear distinction between a community and a rest of the graph. In general, there is a continuum of nested communities whose boundaries are somewhat arbitrary. A community-detection algorithm should therefore return different ``views'', according to the value of some control parameters. At a superficial level, most of our information processing concerns the evaluation of probabilities. When faced with insufficient data or insufficient time for a rational processing, humans have developed algorithms, called heuristic in the cognitive psychology area, that allow us to take decisions in these situations. The modern approach to the study of cognitive heuristics defines them as those \textit{strategies that prevent one from finding out or discovering incorrect answers to problems that are assumed to be in the domain of probability theory}. Basically, the cognitive heuristics program proposed by Goldstein and Gigerenzer suggests to start from fundamental psychological mechanisms in order to design the models of heuristics~\cite{Gigerenzer2002}.  These models have to satisfy the following constraints: (a) \emph{Ecologically rational} (\textit{i.e.}, they exploit structures of information in the environment), (b) Founded in evolutionary psychological capacities such as the memory and the perceptual system, (c) \emph{Fast and frugal}, and simple enough to operate effectively when time, knowledge, and computational power are limited.  We try to implement such human-inspired models in autonomous devices.  We model an ``individual''  as a memory and a set of connections to other individuals, with a simple procedure for filtering information. The information about neighbouring nodes is propagated and elaborated locally over the time as function of the previous meetings. In this way we are able to simulate a process in which the agents, through an alternation of communication and elaboration phases, have their local subjective representation of network. The emerging community knowledge is given by the probability to belong to one or more clusters at the same time. This method, already tested for detecting communities in static networks~\cite{Massaro2012, Bagnoli2012, Guazzini2012}, is now applied to dynamical environments. 
\section{The model}
\label{sec:model}
\begin{figure*}[htb!]
\centering
\subfigure[]
{\includegraphics[width=0.3\columnwidth]{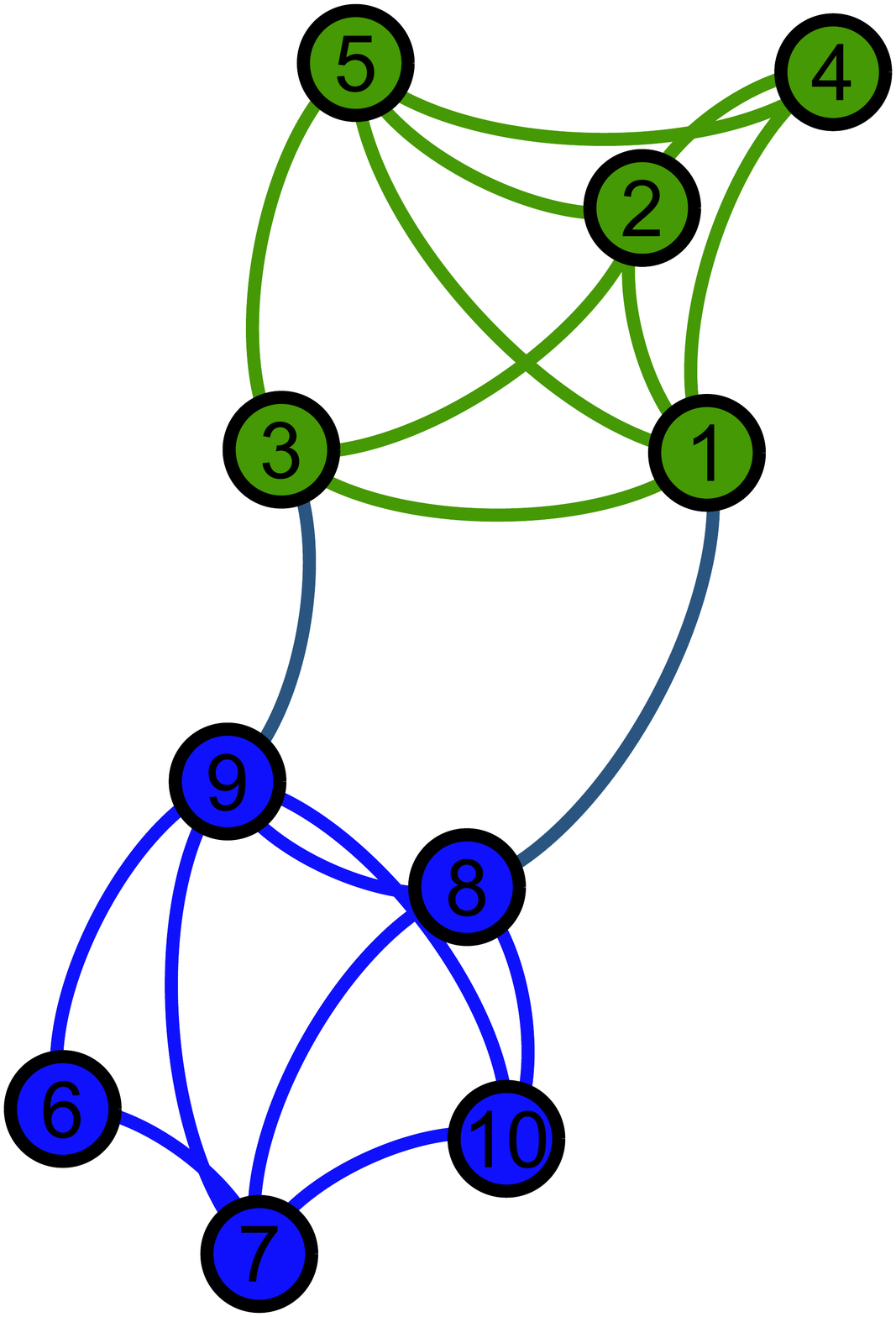}}
\hspace{2mm}
\subfigure[]
{\includegraphics[width=0.4\columnwidth]{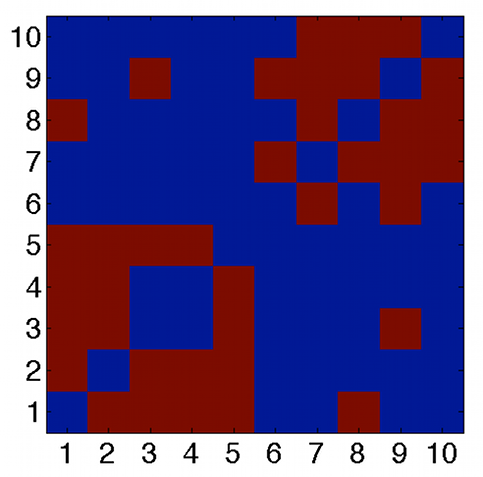}}
\hspace{2mm}
\subfigure[]
{\includegraphics[width=0.455\columnwidth]{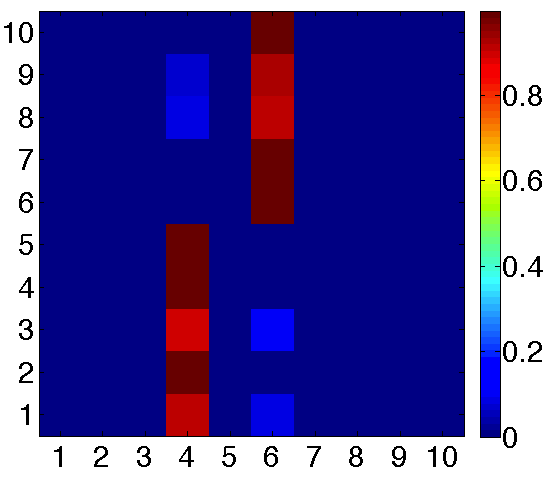}}
\caption{\label{fig:fig1} (a) Network composed by $2$ communities of densely connected vertices ($1-5$ and $6-10$ respectively) with a much lower density of connections between them. (b) Corresponding adjacency matrix: here the red points indicate the presence of a link between nodes $i$ and $j$. (c) Asymptotic configuration of the state matrix $S$, with $m=0.4$ and $\alpha = 1.4$, in which the two principal communities are labelled by \emph{leaves}, nodes $4$ and $6$, that are the nodes with lower connectivity. Moreover it is also possible to detect the overlapping nodes between them, which are the nodes $1$ and $3$ for the first community and the nodes $8$ and $9$ for the second one.  This fact is emphasized by the values of the state matrix where the overlapping nodes have an high probability to belong to their principal community (light red points) but also a low probability to be part of the other one (light blue points). The other nodes have a very high probability to belong only to their principal community (dark red points).}
\end{figure*}
\begin{figure*}[htb!]
\centering
\subfigure[]
{\includegraphics[width=3.5cm]{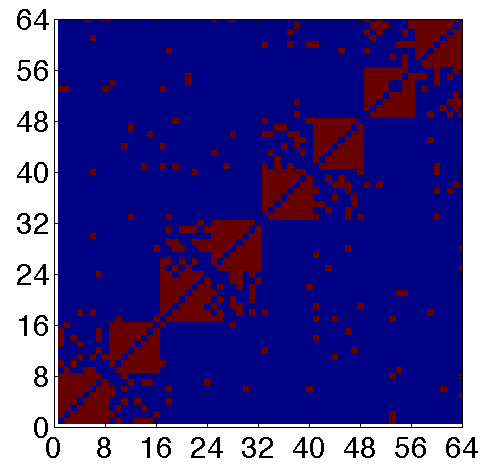}}
\hspace{.1mm}
\subfigure[]
{\includegraphics[width=3.5cm]{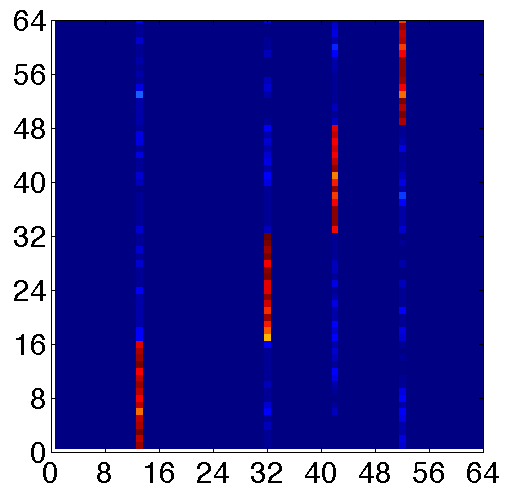}}
\hspace{.1mm}
\subfigure[]
{\includegraphics[width=3.5cm]{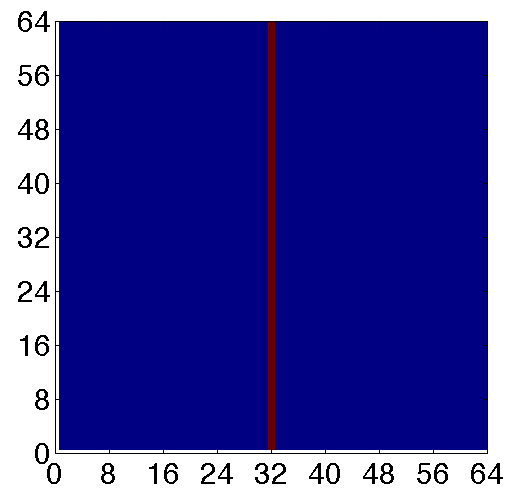}}
\hspace{.1mm}
\subfigure[]
{\includegraphics[height=3.48cm]{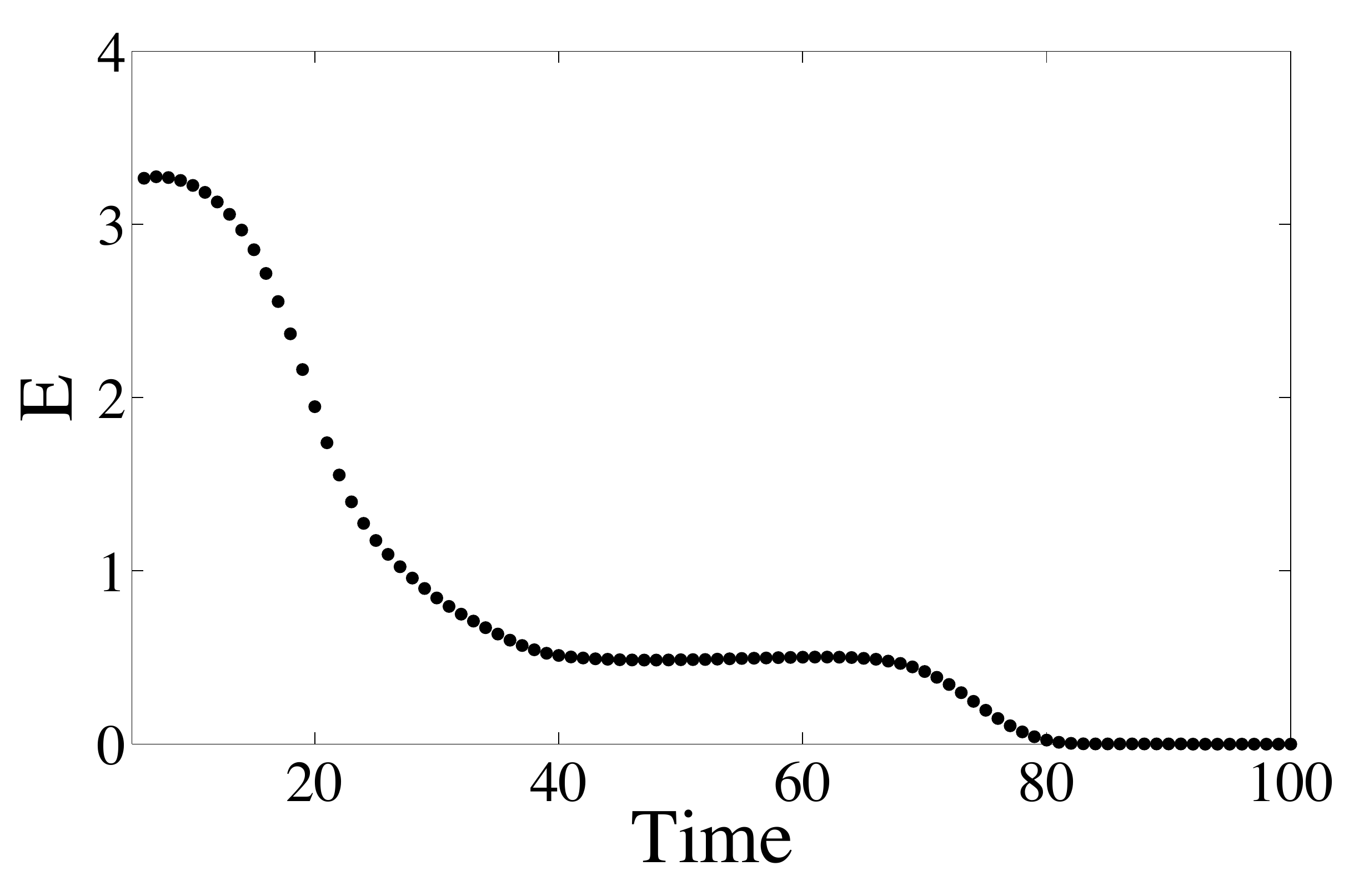}}
\caption{\label{fig:g2} (a) Hierarchical three-level network with $4$ principal
communities by $16$ blocks of $16$ nodes each. (b) Final configuration of state matrix $S$ with $m = 0.2$ and
$\alpha = 1.2$. (c) Final configuration of state matrix $S$ with $m = 0.1$
and $\alpha = 1.2$: the final mono-cluster is identified by the node with lower connectivity in the network. (d) Entropy of information for the whole network during time
regarding the case (c).}
\end{figure*}
\begin{figure}[h!]
\centering
\subfigure[]
{\includegraphics[width=.6\columnwidth]{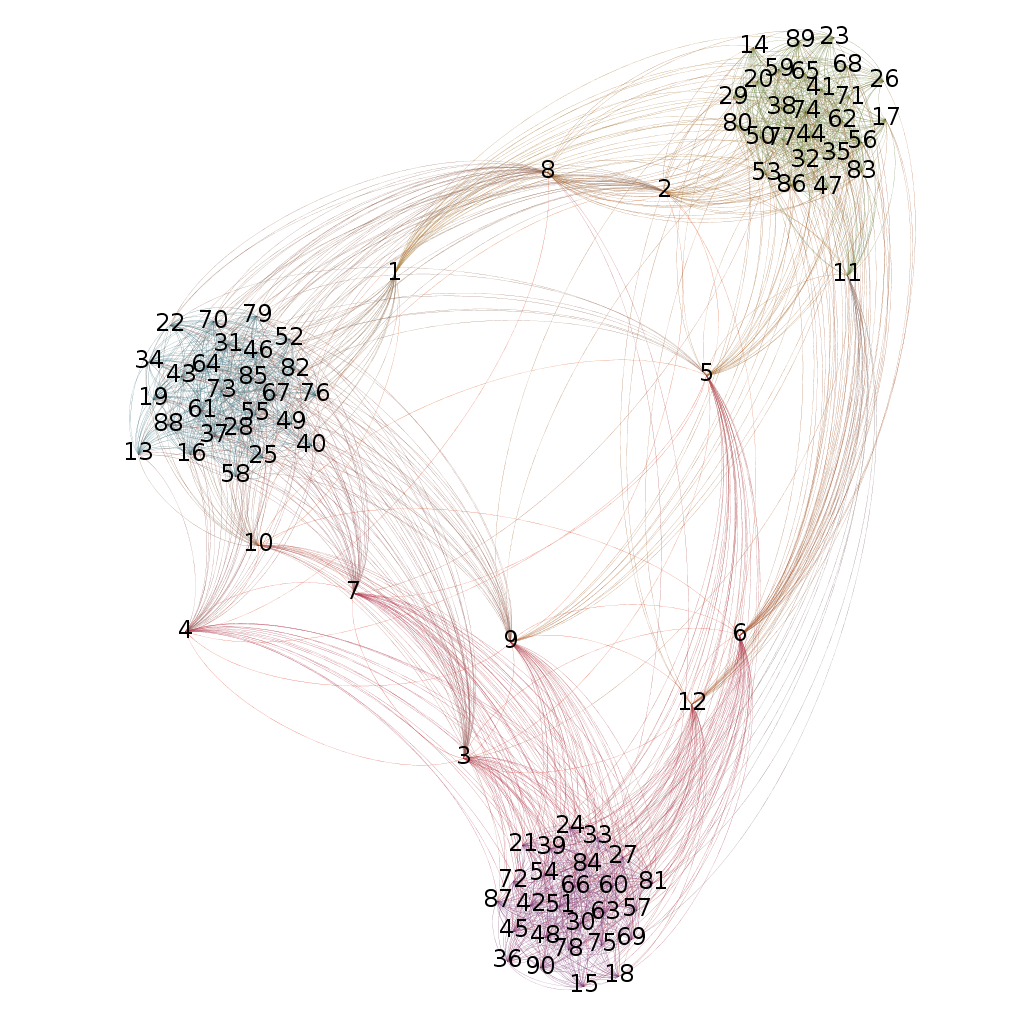}}
\subfigure[]
{\includegraphics[width=.75\columnwidth]{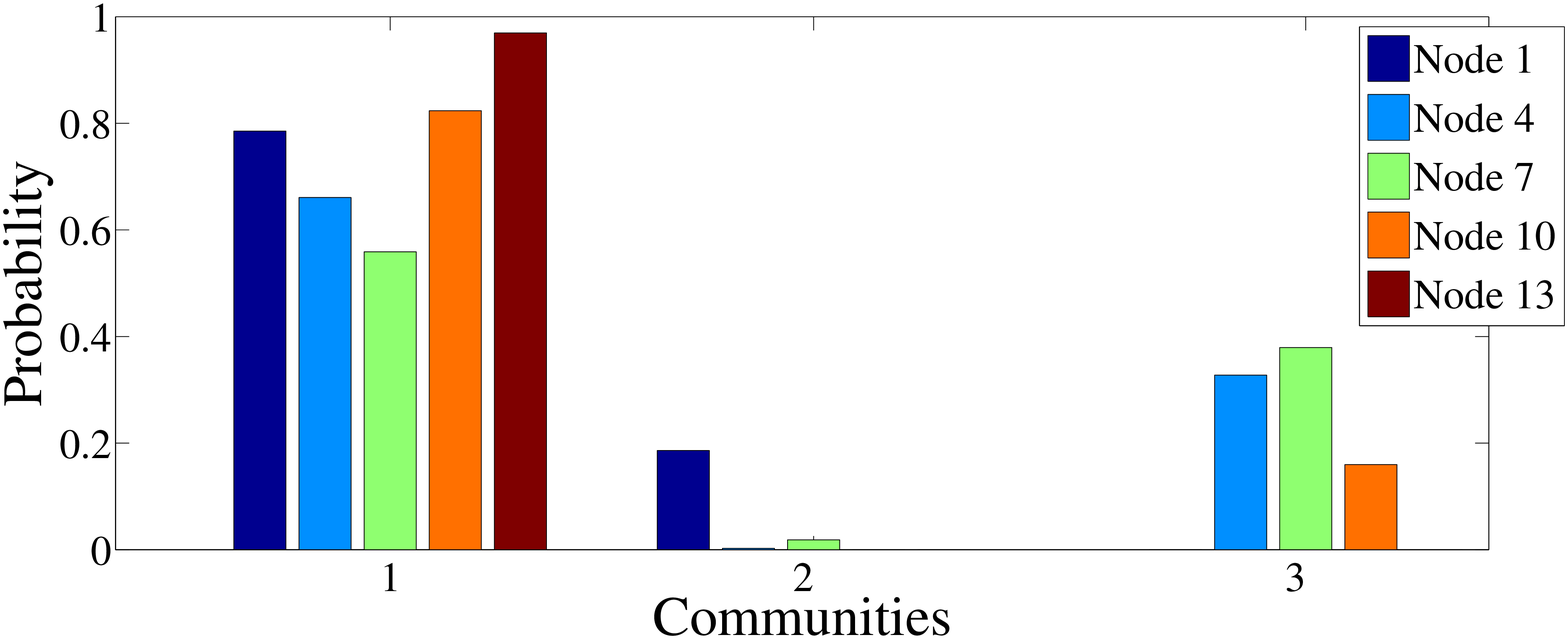}}
\subfigure[]
{\includegraphics[width=.75\columnwidth]{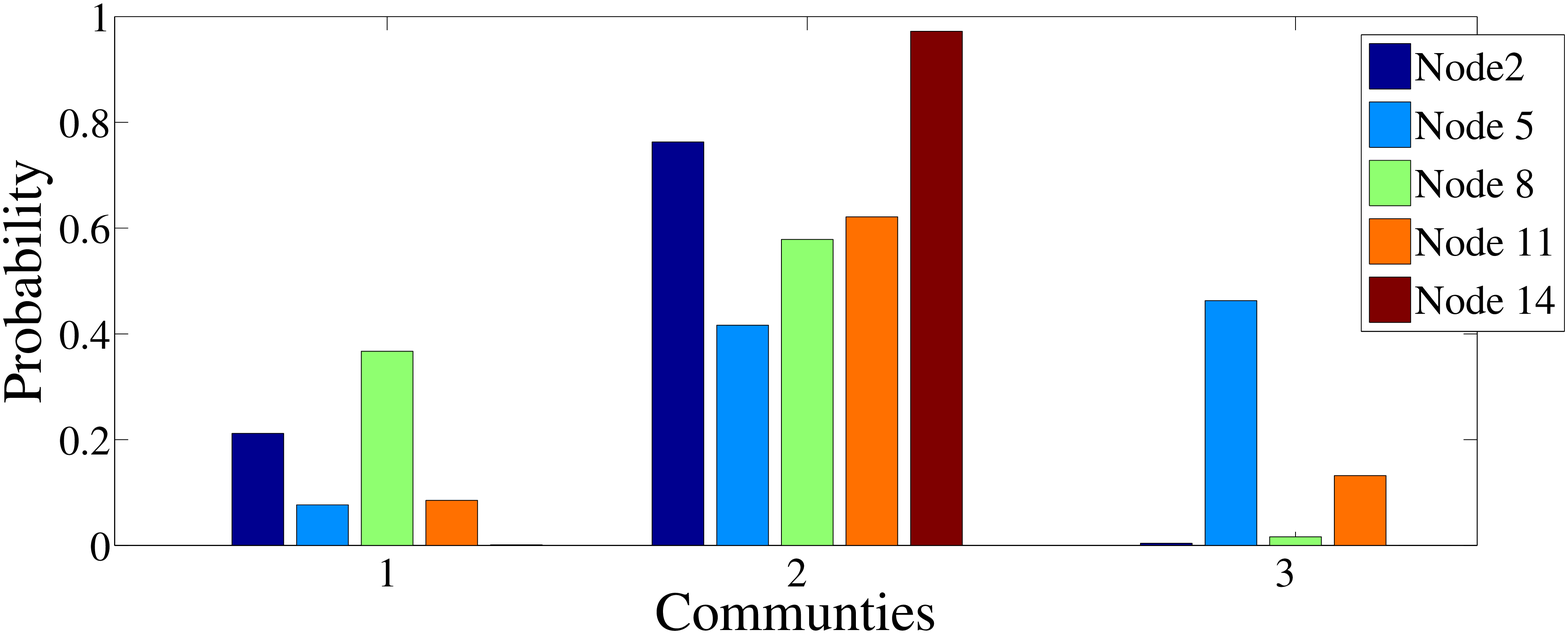}}
\subfigure[]
{\includegraphics[width=.75\columnwidth]{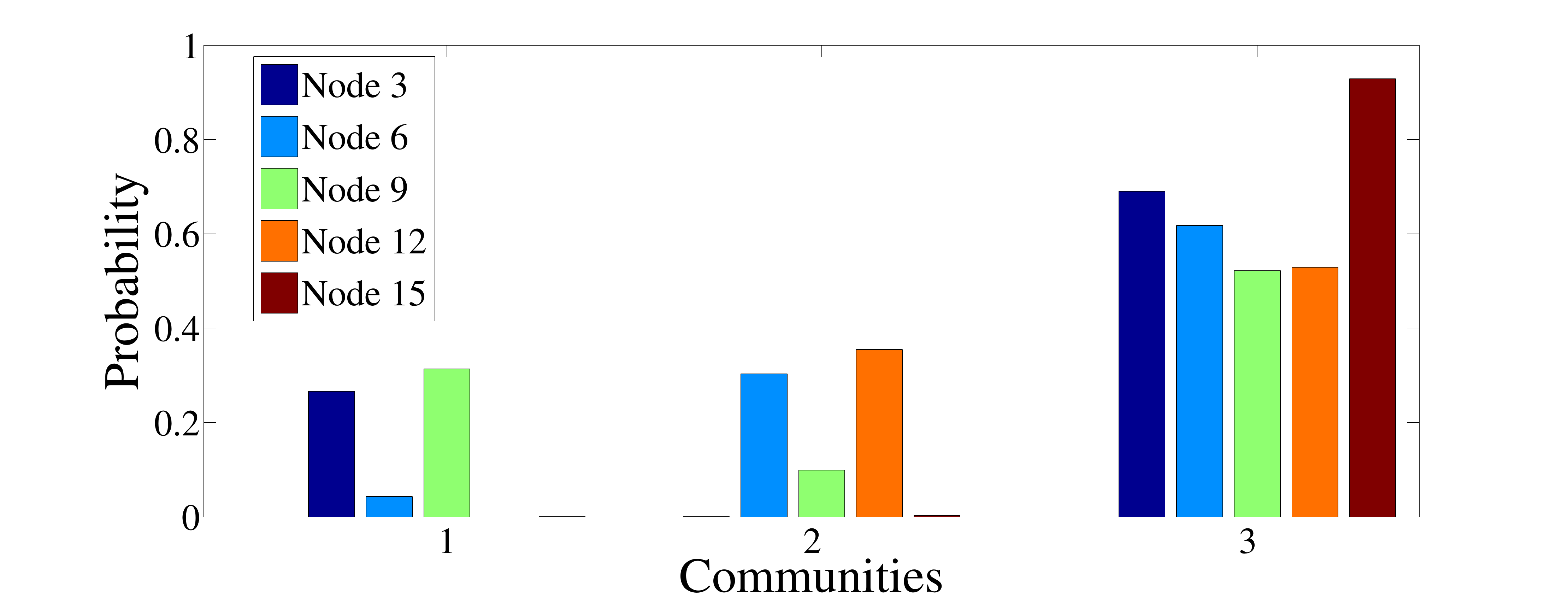}}
\caption{\label{fig:res2} (color online)  Case with $N=90$ nodes, $3$ communities and $N_{tr}=4$ travellers for each community. (a) Community structure of the network revealed by our algorithm where the link represent the encounters between the agents during time while the travellers are the overlapping nodes between the three principal communities.  (b) Probability to belong to the principal communities in the case with $4$ travellers for each community. Local vision of nodes $1, 4, 7, 10 $ and $13$. These nodes are in the same community labelled as $4$. As we can observe the nodes $1,4,7,10$ are the travellers of the community $1$. In fact they belong with a certain probability also to other communities. On the contrary the node $13$ has a very high probability to belong only to its community. }
\end{figure}
Let us first present the static community-detection algorithm derived from the van Dongen's Markov Cluster algorithm (MCL) method~\citep{MCL}. The MCL algorithm simulates a sort of diffusion process over the graph, followed by a pruning phase in which the competition among the links allows to eliminate the weakest ones. In this model the graph is expressed by the correspondent adjacency matrix $A$: specifically, the adjacency matrix of a finite graph \textbf{G} of $N$ vertices is a $N \times N$ matrix where the non-diagonal entry $A_{ij}= 1(0)$ indicates the presence (absence) of a link from the node $i$ to the node $j$, as shown in \figurename~\ref{fig:fig1}(b). The MCL algorithm starts by elaborating the diffusion matrix, which is obtained from the original adjacency matrix by normalizing over rows. In particular the $i-th$ row of $A$ is divided by the connectivity degree $k_i$ of node $i$; then 
\begin{equation}
M_{ij} = \frac{A_{ij}} {k_i},
\end{equation}
where $k_i=\sum_{j=1}^N A_{ij}$. The elaboration is composed by an alternation of expansion and inflation phases. In the expansion phase an integer power $n$ of this matrix -- usually $n=2$ --  is computed,   generating the probability matrix $P$ of an $n$-step random walk. 
Thereafter, in the inflation phase, each element of the probability matrix $P$ is raised to some power $\alpha$
in order to artificially enhance the probability of the random walker of being trapped within a community.  The expansion and the inflation phases are iterated until one
obtains the adjacency matrix of multiple disconnected stars, corresponding to the communities. This method, widely used in bioinformatics, depends strongly on the
choice of the parameter $\alpha$. Its complexity can be partially neglect (or cut off) if, after each step of inflation, only the largest $k$ elements of the resulting matrix are maintained, while the others are set to zero.
Starting from the MCL method, we have developed an algorithm, already described here in Refs.~\citep{Massaro2012, Bagnoli2012, Guazzini2012} and summarised hereafter for the reader's convenience, where  a network of $N$ vertices is represented by its adjacency matrix $A$. The vertices or  nodes are the agents capable to communicate with each other, and each of them has a memory of past encounters (state vector):  each vertex $i$ is characterized by a state vector $S_{ij}(t)$ representing its knowledge about node $j$  at time $t$. We can compactly represent the knowledge of the whole network by a state matrix $S$ ($N \times N$ entries). We suppose that at time $t_0 = 0$ each node knows only itself so  $S_{ij}(0) = 1$ if $i = j$ and $0$ otherwise (\textit{i.e.}, the initial state matrix  the \emph{identity matrix} $\delta_{ij}$). The elaboration of information is modelled as an alternation of communication and elaboration phases. We shall denote by   $S'(t)$ the state matrix after the communication phase and by  $S''(t)= S(t+1)$ the state vector after the elaboration one, \textit{i.e.}, after one whole time step. The information at each node is updated when it encounters another node: two meeting nodes  exchange information about their local view of the network, which is clearly an approximation (due to their partial knowledge) of the real structure of the network. 

\emph{Communication phase}:  In this phase a node passes information about other nodes. His knowledge about other nodes is given by its state vector $S_{ij}$, whose entries are a measure of the relevance of the other nodes. We assume that there is a limitation about the communication time, so that the most relevant informations are communicated with more emphasis (in a real implementation with finite bandwidth, this would imply that the probability of communicating an information about a given node is higher the more relevant that node is).  In order to model this limitation, we normalize the
adjacency matrix on the columns (\textit{i.e.}, we assign at each link the inverse of
the output degree of the incoming node), forming a Markov matrix $M_{ij} = A_{ij}/\sum_k A_{kj}$. We also introduce a memory term $m$ that modulates the evolution of the knowledge:
\begin{equation}
S'_{ij}\left(t + 1 \right) = m S_{ij}\left(t \right) + (1-m) \sum_{k=1}^N M_{ik}S_{kj}(t);
\end{equation}
The parameter $m$ allows us to moderate the \emph{oblivion} effect for which the most recent information is  more important than the old one.

\emph{Elaboration phase}:  The elaboration phase is modelled analogously to the inflation phase in the MCL algorithm:
\begin{equation}
S''_{ij}(t+1) = \frac{S_{ij}^{'\alpha}}{\sum_k S_{ik}^{'\alpha}}.
\end{equation}
This part is also based on the concept of \emph{diffusion and competitive interaction} in network structure introduced by Nicosia et al.~\cite{Nicosia2011}.

Each community is identified by the label of a "characteristic" node (that spontaneously emerge). 
In order to exemplify our method we report the results of the algorithm for the network reported in \figurename~\ref{fig:fig1}(a) represented by the adjacency matrix in \figurename~\ref{fig:fig1}(b) where the red points indicate the presence of a link between nodes $i$ and $j$. This is a network composed by $10$ vertices and two communities $C_1 = 1,2,3,4,5$ and $C_2 = 6,7,8,9,10$. In \figurename~\ref{fig:fig1}(c) we show the image of the final configuration of the state matrix $S$ in which the two communities are labelled by the nodes $4$ and $6$ which are the nodes in the two communities with the lower connectivity degree. Moreover, it is also possible to detect the overlapping nodes between the communities as explained in the figure caption.
The node memory is assumed to be large enough to contain all the pieces of information about other nodes (in a real implementation this should be limited to the most relevant nodes), and the model is characterized by two free parameters: the memory $m$ and the coefficient $\alpha$~\citep{Massaro2012, Bagnoli2012} although it is  possible to let the system automatically tune them as shown in Ref.~\citep{Guazzini2012}. As shown in \figurename~\ref{fig:g2}, the output of the model  depends on the values of parameters. In \figurename~\ref{fig:g2}~(a), an example of a
hierarchical network is presented; the 
three-levels adjacency matrix is composed by $8$ blocks of 8 nodes (first-level communities), grouped in  $4$  second-level communities of $2$
blocks, with a link probability that is respectively of $0.98$ inside
blocks, $0.3$  among blocks in the the second-level communities, and $0.03$ among the rest.  The red points indicate the presence of a link between  node $i$ and  node $j$, $A_{ij} = 1 $. 
In \figurename~\ref{fig:g2} (b), the asymptotic configuration of the
matrix $S$ is shown using $m = 0.2$ and $\alpha = 1.2$, while in
\figurename~\ref{fig:g2} (c) it is computed using  $m = 0.1$ and $\alpha = 1.2$. It can be noticed that in the first case the algorithm discovers the four second-level communities, and the second case all nodes belong to the same community. 
In order to present the data in a compact way, let us introduce 
the information entropy $E$, defined
as
\begin{equation}
E^{(S)} = - \sum_i P_i^{(S)} \log(P_i^{(S)})
\end{equation}
where $P_i^{(S)} = \sum_i S_{ij}$. 
The entropy $E$ reaches the maximum  for the flat distribution, where each
node knows only itself, and reaches a minimum (zero) when all nodes know the
same label (i.e. all state vectors are the same and contain just one element
different from zero). It is possible to follow the evolution of the global knowledge by plotting the value of the entropy $E(t)$ during time, as shown  in~\figurename~\ref{fig:g2} (d) corresponding to the parameters of case (c). Although the final state is that of minimum entropy (only one label), it is possible to see that the network identifies during time the different levels of the hierarchical structures, showing them as  plateaus in the entropy plot.

It is  possible to apply this method to dynamical networks. 
In this case the adjacency matrix $A_{ij}(t)$ changes in time, due to  the displacement of  agents. At each time step each node saves its local vision of the network in order to have the right view during  time, as we show in the next Section.

\section{Results}
\subsection{Simulated environment}
We apply our algorithm to the case of nodes that move as in one of the reference  models in the opportunistic networking literature, the HCMM \cite{Boldrini:2010fk}, already used in several works to evaluate the performance of data forwarding and  dissemination  for OppNets~\cite{Allen2012, Picu2012}. This allows us to show that our algorithm can be used to dynamically detect the structure of communities of users in mobile social networking environments. 
 Mobility traces generated by HCMM incorporate temporal, social and spacial notions in order to obtain a proper representation of the real user movements. More precisely, nodes move in an area of $1000\ \mathrm{m}^2$ divided in a  $6 \times 6$ grid where a single grid's cell represent a physical location that corresponds to a community. In this synthetic scenario, communities are placed far from each other so  to avoid any border effect, \textit{e.g.}, involuntary communication between groups. In each community we place two kinds of moving nodes: travellers and non-travellers. Non-travellers roam only inside their community, while travellers, from time to time, use to visit other social communities different from the one they belong to.  In this context, the only way to exchange information is through nodes mobility, and travellers play an important role because they are the unique bridge between communities. We only use proximity information, so edges correspond to contacts. We do not use other social information.

 In our experimental set-up, we consider a network of N = 90 nodes, divided in  3 separated communities and we  study the performance of the algorithm by incrementally increasing the number of travellers for each  community. We want to evaluate the average discovery time of the underlying community structure together with the goodness of the detection itself. Indeed, by increasing the number of travellers  the information flow from one community to another also increases, but the actual community boundaries becomes less defined, making the community detection problem more and more challenging.
 
For simplicity, we used the same time step for the alternating computation and the user mobility, but clearly in a real world the elaboration phase would be much faster than the mobility one.  
 
 The detailed scenario configuration can be found in Table \ref{tab.params}.
 
 \begin{table*}[t]
\centering
\caption{Detailed scenario configuration\label{tab.params} }
\begin{tabular}{|c|c|}
\hline
\textbf{Paramenter} & \textbf{Value} \\
\hline
Node speed & Uniform in $[1,1.86m/s]$\\
Transmission range & $20m$\\
Simulation Area & $1000 \times 1000m$ \\
Number of cells & $6 \times 6$ \\ 
Number of nodes & $90$ \\
Number of communities & $3$\\
Number of travellers per community &  $3,4,7,13$\\
Simulation time & $50000s$\\
\hline
\end{tabular}
\end{table*}

\subsection{Performance evaluation}

\begin{figure}[htb!]
\centering
\subfigure[]
{\includegraphics[width=0.82\columnwidth]{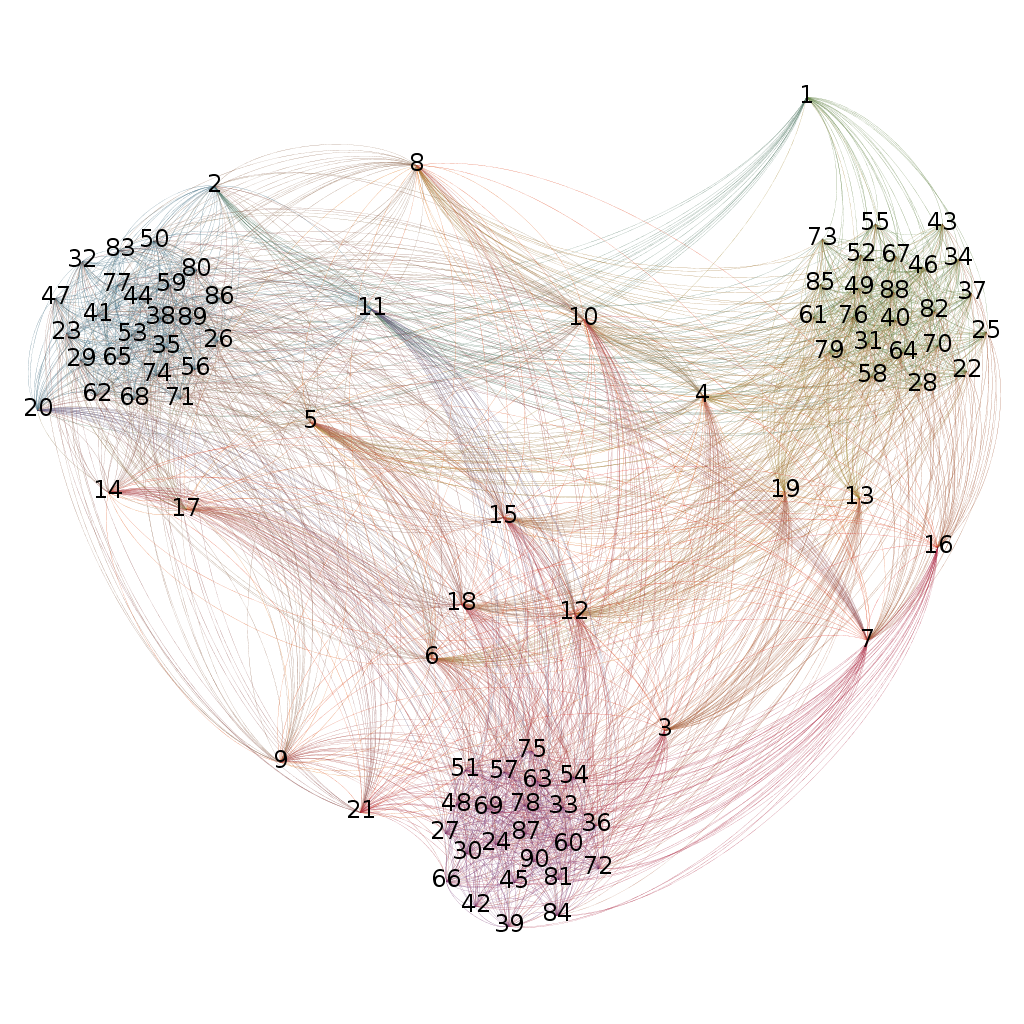}}
\subfigure[]
{\includegraphics[width=0.82\columnwidth]{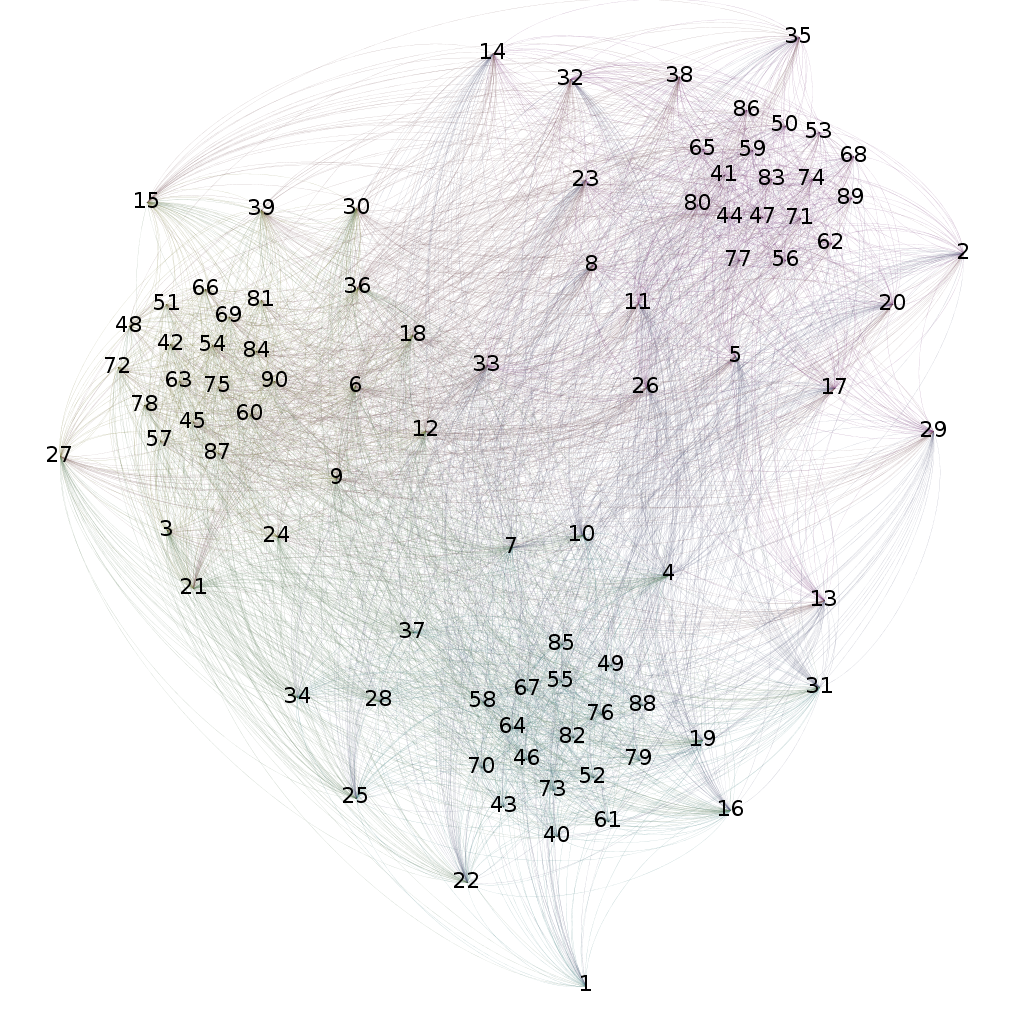}}
\caption{\label{fig:res4} (color online) Final community structures detected by the algorithm considering (a) $N_{tr}=7$ and (b) $N_{tr}= 13$ travellers respectively.}
\end{figure}

\begin{figure}[htb!]
\centering
\includegraphics[width=1\columnwidth]{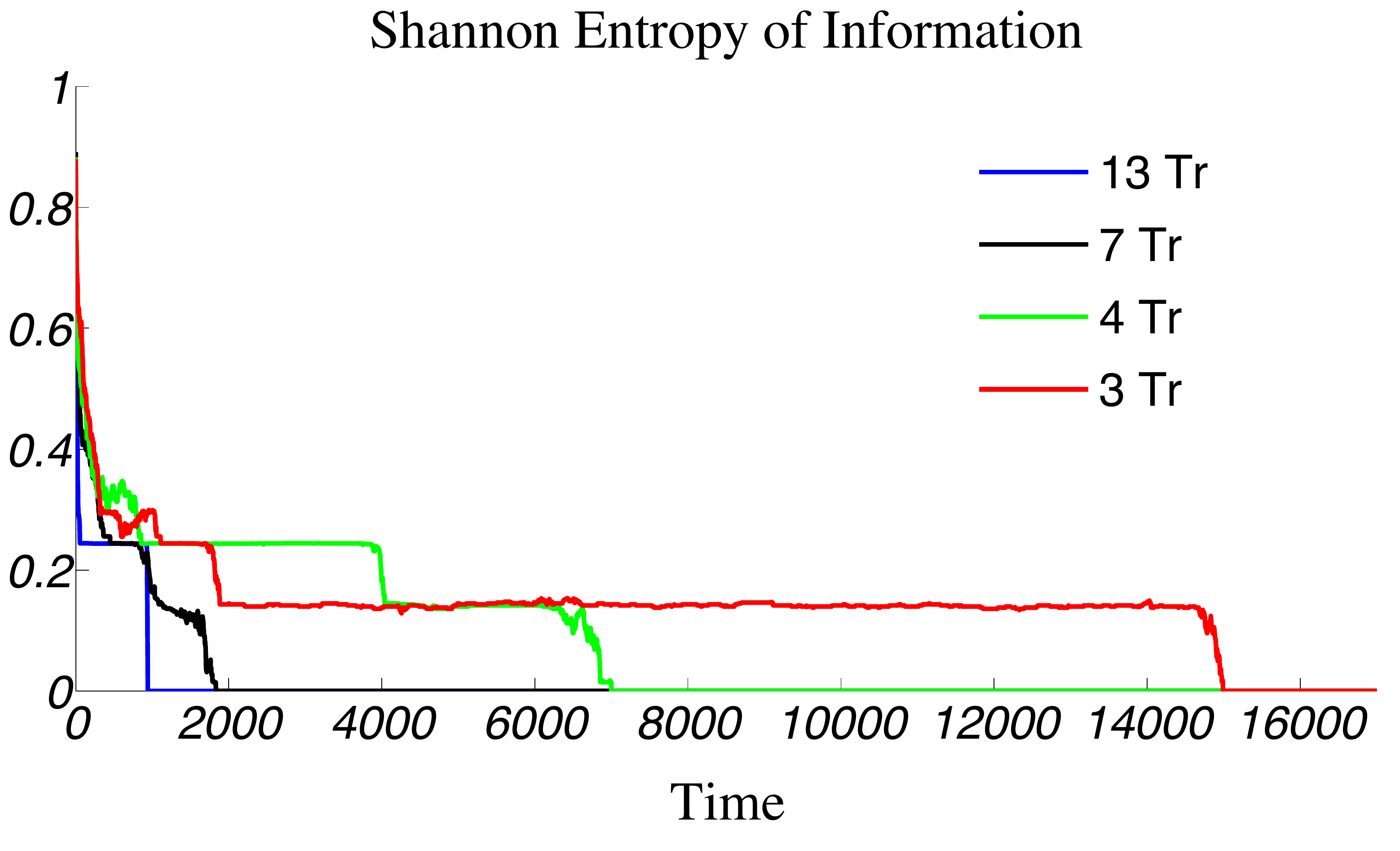}
\caption{\label{fig:res3} (color online) Information entropy $E(t)$ during time for different scenarios with $N_{tr}=13$, $N_{tr}=7$, $N_{tr}=4$ and $N_{tr}=3$ travellers.}
\end{figure}

The results of the algorithm  with $4$ travellers for each community is shown in \figurename~\ref{fig:res2}. In \figurename~\ref{fig:res2}(a) we show the snapshot of the community structure revealed by our algorithm. We can observe the $3$ principal clusters but also the overlapping nodes between the communities  that correspond to the travellers. The state matrix $S$  is the probability for a node to belong to a certain community: this data is reported in \figurename~\ref{fig:res2}(b)-(c)-(d) where the bars of the histogram corresponds to the probability for the nodes to belong to the a given community. For instance, looking at \figurename~\ref{fig:res2}(b) we can observe that nodes $1,4,7,10$ and $13$ have an high probability to belong to  community $1$ but the first four nodes have also a little probability to belong  to other communities. In fact, node $1$ (blue bar in \figurename~\ref{fig:res2}(b)) is a member of  community $1$ with $p \sim 0.78$ and of the community $2$ with $p \sim 0.22$ because it is a traveller  between the two communities. While the node $13$ has a probability $p \sim 1$ to belong to the community $1$: in this way each node  is aware of its role inside its community. 

In \figurename~\ref{fig:res4}(a)-(b) we report the snapshots of the final community structure detected by our algorithm considering $7$ and $13$ travellers, respectively: also here the algorithm is able to detect not only the three principal clusters but also the travellers as the overlapping nodes between the communities.

In \figurename~\ref{fig:res3} we show the different plots of the information entropy for different cases considering different number of travellers. Here we can  not only observe the three plateaus corresponding to three principal clusters, but also the converging times for reaching the final state. By increasing the number of travellers, the time for reaching the asymptotic state decrease. The convergence time can be used therefore as an indicator of the performances of the detection and as a measure of the ``boundary size'' of the community.

\begin{figure}[htb!]
\centering
\includegraphics[width=.8\columnwidth]{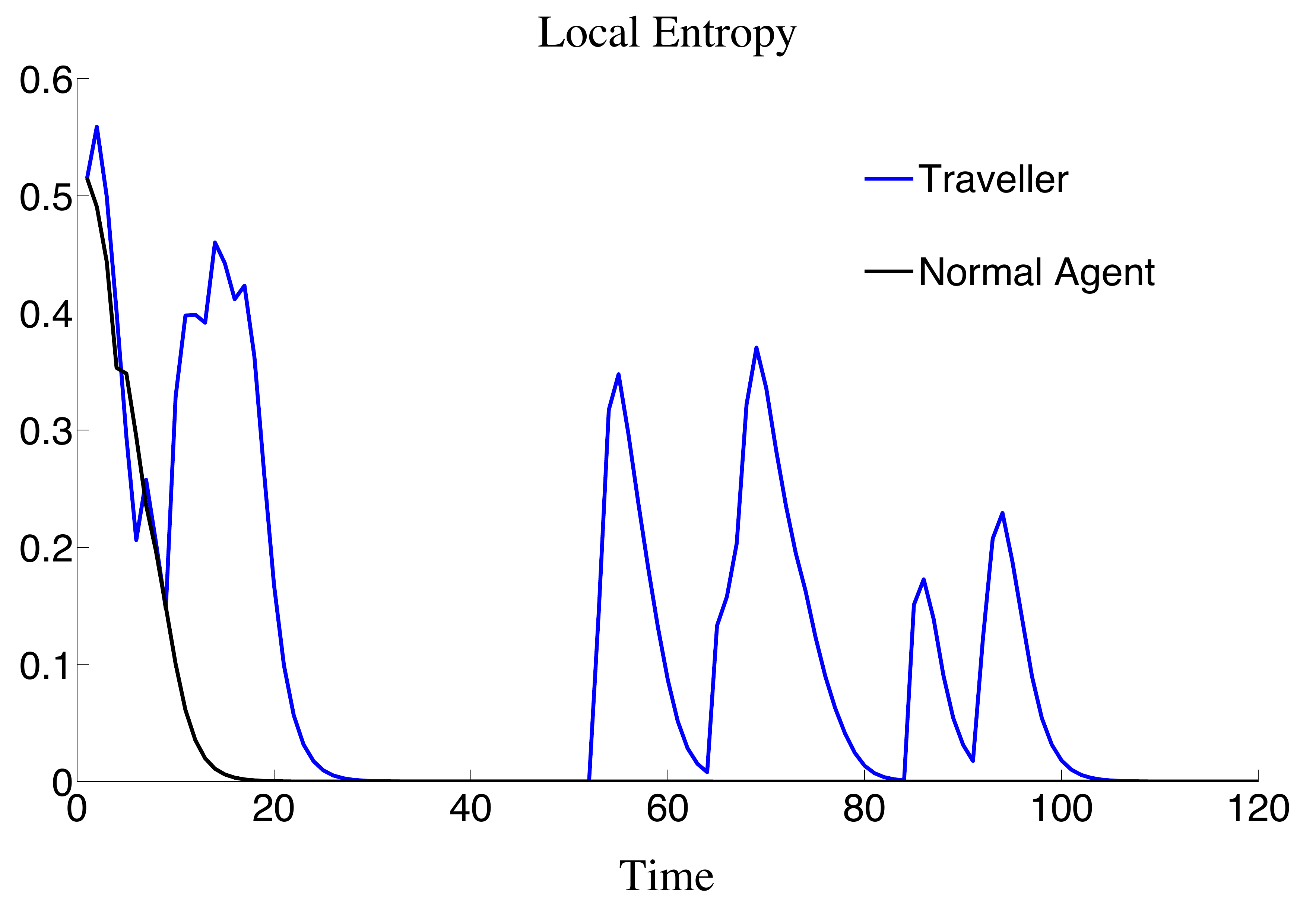}
\caption{\label{fig:res5} (color online) Comparison between the local entropy of a traveller (jumping blue line) and a normal agent (black line). The peaks correspond to the switched to other communities.}
\end{figure}

Finally, in \figurename~\ref{fig:res5} we report the local entropy for a traveller (black line) and for a normal agent (blue line) during time. The local entropy $E^i$ is simply define as $E^{(i)} = -\sum_j^N S_{ji}\log S_{ji}$ and represents the knowledge of the single node about the surrounding world. While the knowledge of a normal agent quickly relaxes to a stationary value, that of travellers exhibits jumps when the agent switches to other communities.

\section{Conclusion and future work}
In this paper, we proposed a local cognitive-inspired community detection algorithm for opportunistic networking environments. Given the growing interactions between mobile devices and humans we focused our attention on the importance of the spreading and elaboration of the information which has a crucial role in CPW \cite{Conti20122}. We evaluated it on different synthetic human mobility scenarios and we found that our method is capable to detect not only the right communities from an individual viewpoint but also to spontaneously reveal the role of each nodes inside the network (\emph{travellers} and normal agents) providing a natural ``scanning'' of the various clustering levels.  In the future, we would like to evaluate the scaling of our algorithms with the system size and  apply it to more realistic scenarios. In particular we plan to compare our algorithm  with others targeted to pocket switched networks (that use also global information)~\cite{Hui, williams}. We would also like  to combine the geographic proximity with additional social information so as to better catch the complex association between the   real and the virtual world.

\section*{Acknowledgments}
This work is partly funded by the EC under the FET-AWARENESS RECOGNITION Project (FP7-257756) and by the EIT ICT Labs Emergent Social Mobility Project.

\bibliographystyle{unsrt}
\bibliography{biblio_sca}


\end{document}